\documentclass[aps,prl,twocolumn,floatfix, showpacs]{revtex4}
\usepackage{graphicx}
\usepackage{dcolumn}
\usepackage{bm}
\usepackage[]{natbib}

\newcommand{\be}{\begin{equation}}
\newcommand{\ee}{\end{equation}}

\begin{document}

\title{Streaks to Rings to Vortex Grids: Generic Patterns in Transient Convective Spin-Up }

\author{J.-Q. Zhong,$^1$ M.~D. Patterson$^{1,4}$~J.~S. Wettlaufer$^{1,2,3}$}
 
\affiliation{$^1$Department of Geology \& Geophysics, $^2$Department of Physics and \\
$^3$Program in Applied Mathematics, Yale University, New Haven, CT 06520-8109, USA}
\affiliation{$^4$Faculty of Engineering \& Design, University of Bath, Bath, BA2 7AY, UK}

\date{\today}
 
\begin{abstract}

We observe the transient formation of a ringed pattern state during spin-up of an evaporating fluid on a time scale of order a few Ekman spin-up times.  The ringed state is probed using infrared thermometry and particle image velocimetry and it is demonstrated to be a consequence of the transient balance between Coriolis and viscous forces which dominate inertia, each of which are extracted from the measured velocity field.  The breakdown of the ringed state is quantified in terms of the antiphasing of these force components which drives a Kelvin-Helmholtz instability and we show that the resulting vortex grid spacing scales with the ring wavelength.  This is the fundamental route to quasi-two dimensional turbulent vortex flow and thus may have implications in astrophysics and geophysics wherein rotating convection is ubiquitous. 

\end{abstract}


\maketitle

\begin{figure}
\includegraphics[width=3.15in]{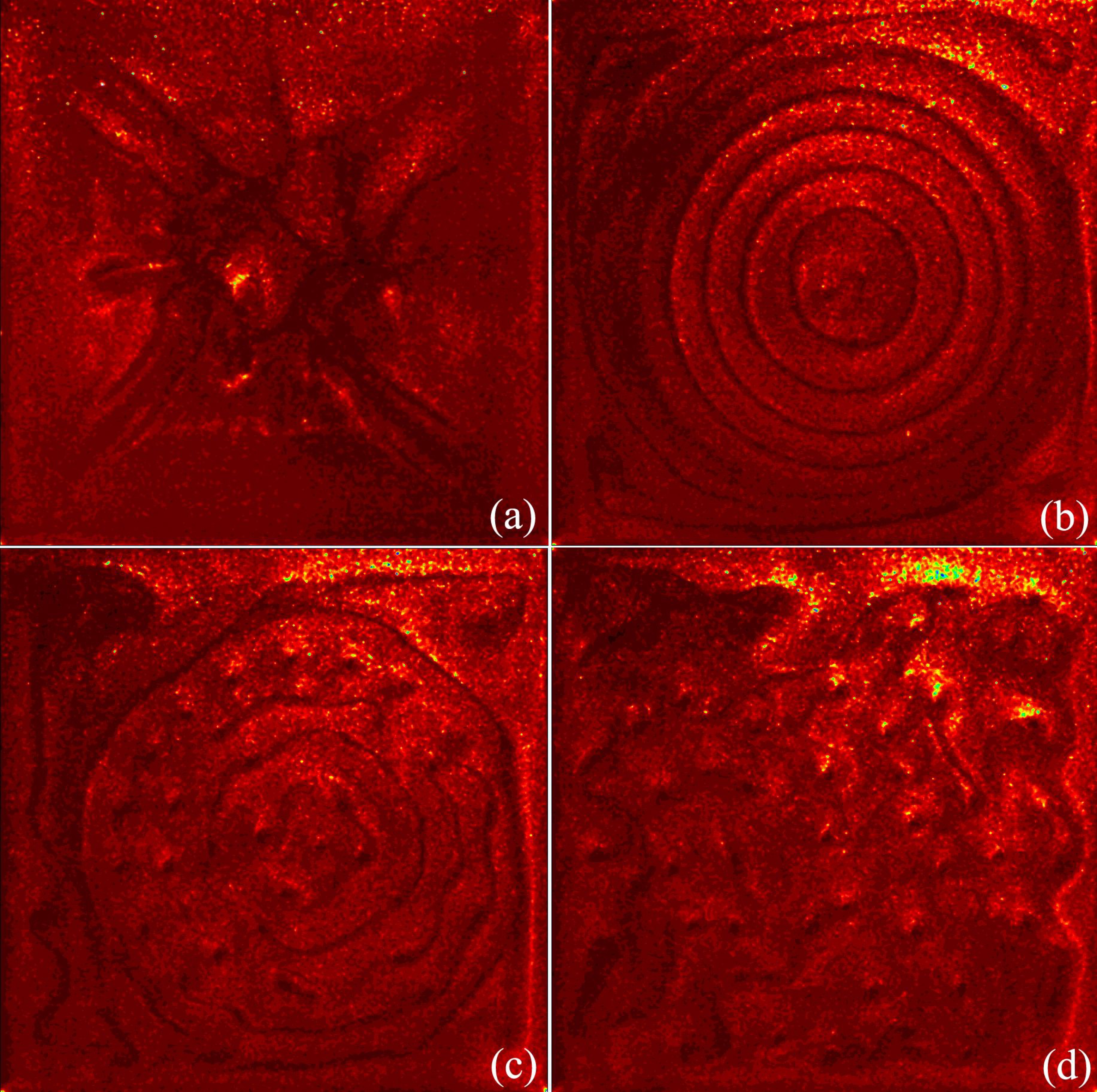}
\caption{A sequence of  particle-number-density images showing the flow field at a horizontal level 0.6 cm below the free surface of a volume of water $H$ = 11.4 cm deep with a 22.9 $\times$ 22.9 cm  cross section. Unless otherwise specified dimensionless time is $t \equiv {t^\prime}/\tau_E$, with time = $t^\prime$. (a) $t=0$ just before the initiation of rotation. (b) $t=2.6$, the flow field was axisymmetric consisting of regular concentric rings. (c) $t=3.9$, wavy billows appeared upon the breakdown of the rings. (d) $t=5.0$ the vortex grid state. In this experiment $Ro=1.4{\times}10^{-3}$ and $\tau_E=167$ s.  The movie is online \cite{Movie}.
}
\label{Figure1}
\end{figure}

The impulsive rotation, or change in angular velocity $\Omega$, of bounded fluid described by the Navier-Stokes equations results in the propagation of stresses into the interior.  Despite the ubiquity of the process, it remains sufficiently complex to insure its centrality as a long standing and  fundamental problem in fluid mechanics \cite{Bondi:48, Harvey:63}.  Fluids are often heated or cooled at their bounding surfaces and whether they are homogeneous or stratified, there are common conundrums that influence a range of disciplines from astrogeophysical flows  to engineering phenomena~\cite[e.g.,][]{EAS:1968, GV:70,Duck:2001,Flor:2004,WCV07, Chandrabook}.   Here, we examine pattern formation during transient spin-up experiments where buoyancy forcing is supplied by evaporative cooling from the top surface of a layer of water of depth $H$ that is probed using infrared thermometry and particle imaging velocimetry (PIV).  It is found that this localized surface cooling drives convection that evolves into a radially symmetric ringed, or ``bulls-eye'',  pattern near the free surface (Fig. \ref{Figure1}b) itself decaying vertically creating a ``bowl'' shaped region (Fig. \ref{Figure5}).  We extract velocity fields to uncover the origin of the transient, symmetric and depth-dependent ring pattern.  It is demonstrated that this results from a temporary balance between the Coriolis and viscous forces which dominate inertia during an Ekman spin up time $\tau_E \equiv \sqrt{H^2/{\nu \Omega}}$
of order unity, where $\nu$ is the kinematic viscosity.  Finally, we find that beyond this transient state,  the ring pattern breaks down to the canonical quasi-two dimensional vortex grid state  \cite{BG:1986, VE:1998} in a manner reminiscent of a Kelvin-Helmholtz instability (Fig. \ref{Figure1}c).  It is found that the number density of vortex cores $N$ is inversely proportional to the square of the ring wavelength $\lambda$, itself decreasing with $\Omega$. 

\begin{figure}
\includegraphics[width=3.15in]{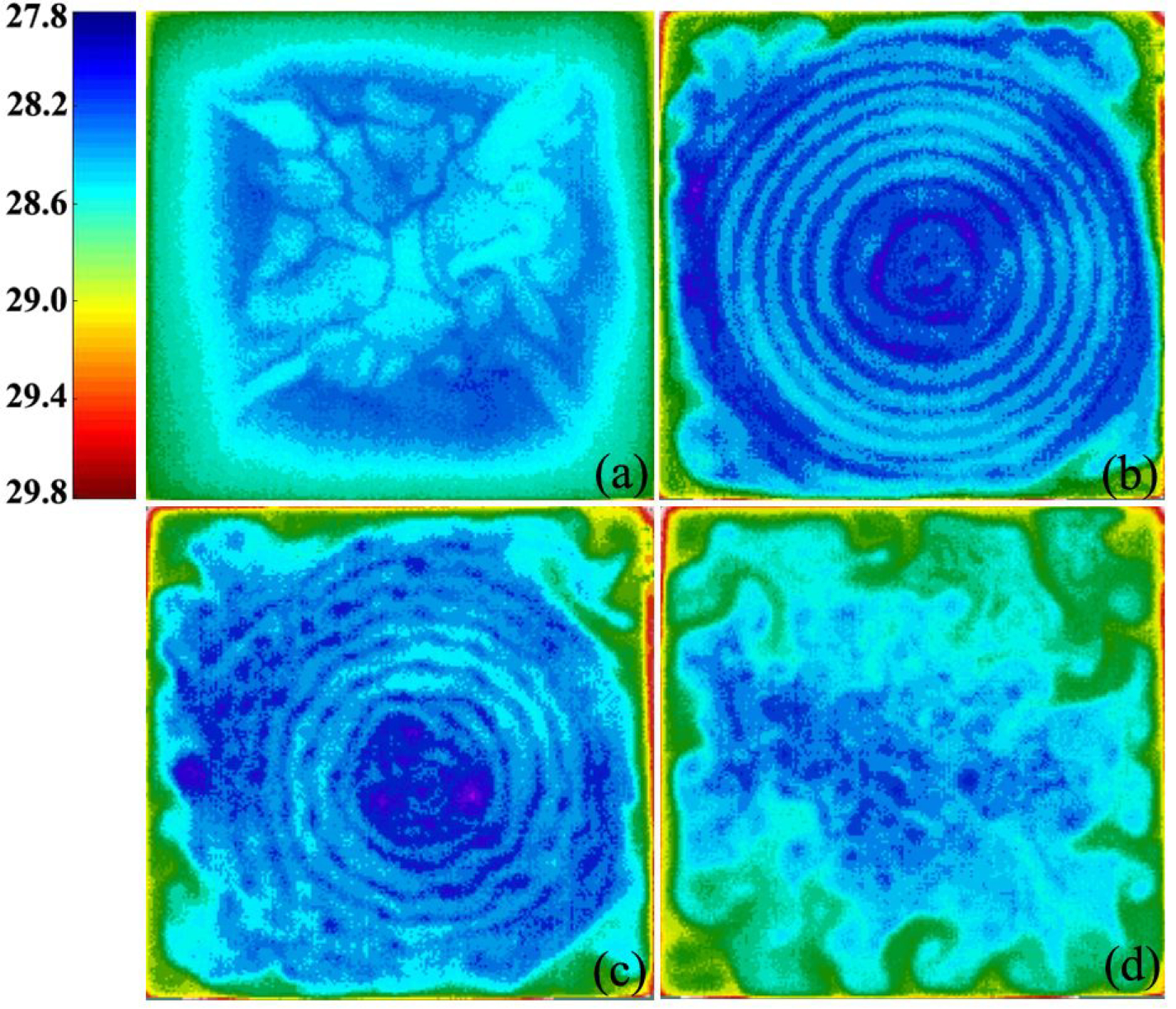}
\caption{Thermal images of the surface taken at four stages during the spin-up process. In the experiment $Ro=5{\times}10^{-4}$, the rotation period is 5.5 s and hence ${\tau_E}$ = 119 s. The images were taken at (a) $t = 0$, (b) $t = 2.9$, (c) $t = 3.7$, (d) $t = 8.7$. The color bar is the temperature scale in $^o$C.}
\label{Figure2}
\end{figure}

The patterns that emerge in rotating convection display the competition between Coriolis and viscous forces across a range of space and time scales \cite[e.g.,][]{Chandrabook, Duck:2001}.  
A signature process in rotating convection contrasts the implications of Taylor-Proudman theorem, which forbids vertical motion for an inviscid constant density rotating fluid, with the buoyancy forcing, which can drive narrow regions of weak convection in fluids with finite viscosity.  Thus, the
formation of vortices or sheets in the interior of a rotating fluid subject to buoyancy forcing can be characterized by the flux Rayleigh number $Ra$ describing the strength of buoyancy transfer in the fluid,  the Rossby number $Ro$ capturing the relative influence of
buoyancy forcing to rotation and the Taylor number $Ta$ characterizing the relative influence of rotation to viscous forces: 
\be
Ra=\frac{BH^4}{\kappa^2\nu},~Ro=\sqrt{\frac{B}{\Omega^3H^2}}, ~Ta=\left(\frac{2 \Omega H^2}{\nu}\right)^2 ,
\ee 
where $B$ is the buoyancy flux and $\kappa$ is the thermal diffusivity of the fluid (in our case water). 
It is well known that in rotating and non-rotating convection \cite[e.g.,][]{Chandrabook}  the critical Rayleigh number ($Ra_c$) for the onset of convection depends on the nature of the thermal boundary conditions (fixed temperature, fixed flux, or mixed) and that in the former case whether both bounding surfaces are free, rigid or one is free and one is rigid; $Ra_c$ converges to a large Taylor number scaling of $Ra_c \rightarrow \mbox{constant} ~Ta^{2/3}$ \cite{Chandrabook}.  Far beyond critical, a particularly compelling example of convective pattern formation is the transient ring state observed during spin-up.  The present study, in combination with previous work \cite{BG:1986, VE:1998}, confirms that this transient state, and its breakdown into a vortex grid is generic in that it occurs under all thermal and mechanical boundary conditions described above, as well as in both cylindrical and square cross section cells.   Our method quantifies the underlying causes for the evolution of the entire process.   

The square cross-section cell containing deionized water initially at rest is set into rotation about its vertical central axis to values of $\Omega$ from 0.1 to 1.0 rad s$^{-1}$.  The bottom and side walls are insulated and maintained at the same temperature $T_w=30~{\pm}~0.1^o$C and the relative humidity above the fluid is held constant at 35$\%~{\pm}~1\%$ during the course of an experiment.   Thus, evaporation induced a constant heat flux $Q$ = 56.0 W m$^{-2}$ driving a buoyancy flux $B$ =  $3.7 \times 10^{-8}$m$^2$s$^{-3}$, so that $Ra$ = $3.3 \times 10^8 \gg Ra_c$, $5 \times 10^{-4} \le Ro \le 10^{-2}$ and $2 \times10^7 \le Ta \le 1.4 \times 10^9$, thereby insuring a rotationally dominated regime of flow.

The number density of seed particles in the flow is captured using a grid-based PIV system \cite{PIV} to image the flow structure as a function of time. The velocity data were calculated in a 1000 $\times$1000 grid with a spatial resolution of 0.2 mm/pixel.  For $\Omega$ = 0, Fig. \ref{Figure1}(a) shows a random network of thin sheet-like downwelling cold streamers typical of evaporatively-induced turbulent convection \cite{BERG:1966}. Shortly after rotation begins, these downwelling plume structures are sharply modified and organized into concentric rings. The rings form first  at the center, then sequentially grow to cover most of the horizontal plane by $t = 2.6$ rotating counterclockwise with uniform azimuthal velocity in a quasi-stationary state  (Fig.\ref{Figure1}b).  
Kelvin-Helmholtz billows appear on the rings at $t=3.9$, grow and roll up into counterclockwise vortices (Fig.\ref{Figure1}c).  Finally, at $t=5.0$, the azimuthal symmetry of these vortices is lost as they assemble into the regular vortex grid seen in Fig.\ref{Figure1}(d).  At this stage we imaged the vertical fields to find classical rotating convective columnar structure \cite{BG:1986, BG:1990, Marshall:1999, VE:2002, Sprague:2006, WW:2008}.  

Infrared thermometry confirms the hypothesis that the basic dynamics of ring formation and breakdown is the same with free surface evaporative cooling and bottom heating \cite{BG:1986},  Rayleigh-B\'{e}nard \cite{VE:1998} boundary conditions, and free surface evaporative cooling with insulated bottom boundary conditions as studied here.  The surface temperature maps (Fig. \ref{Figure2}) reveal the full evolution of the thermal forcing and associated pattern formation, and a positive correlation between thermal fluctuations and vertical velocity \cite{Shang:2003, Brown:2005}. The evolution of the patterns seen in Fig.\ref{Figure2}(a-d) are to be compared with those in Fig.\ref{Figure1}(a-d). Although these are two separate runs, the evolution is essentially the same; the overall temperature drop from the wall to the center is $\Delta{T} \approx$ 2.0$^o$C and is the same at different stages of spin-up.  The width of the vertical thermal boundary layer at the wall is $\delta_T \approx$ 10 mm, evolving slowly in time.

Fig.(\ref{Figure2})b is the map for the maximal ring structure having a radial temperature amplitude of oscillation $\delta{T} = 0.2^o$C~$= 0.1\Delta{T}$ and a minimal $\delta_T$.  The radial oscillation between descending (dark) and ascending (light) fluid is clearly seen, finally merging with the thermal boundary layers at the walls.  When the Kelvin-Helmholtz instability is initiated in Fig.(\ref{Figure2})c  the ringed-shape thermal plumes break down into vortices, the centers of which contain sinking fluid.

\begin{figure}
\includegraphics[width=3.15in]{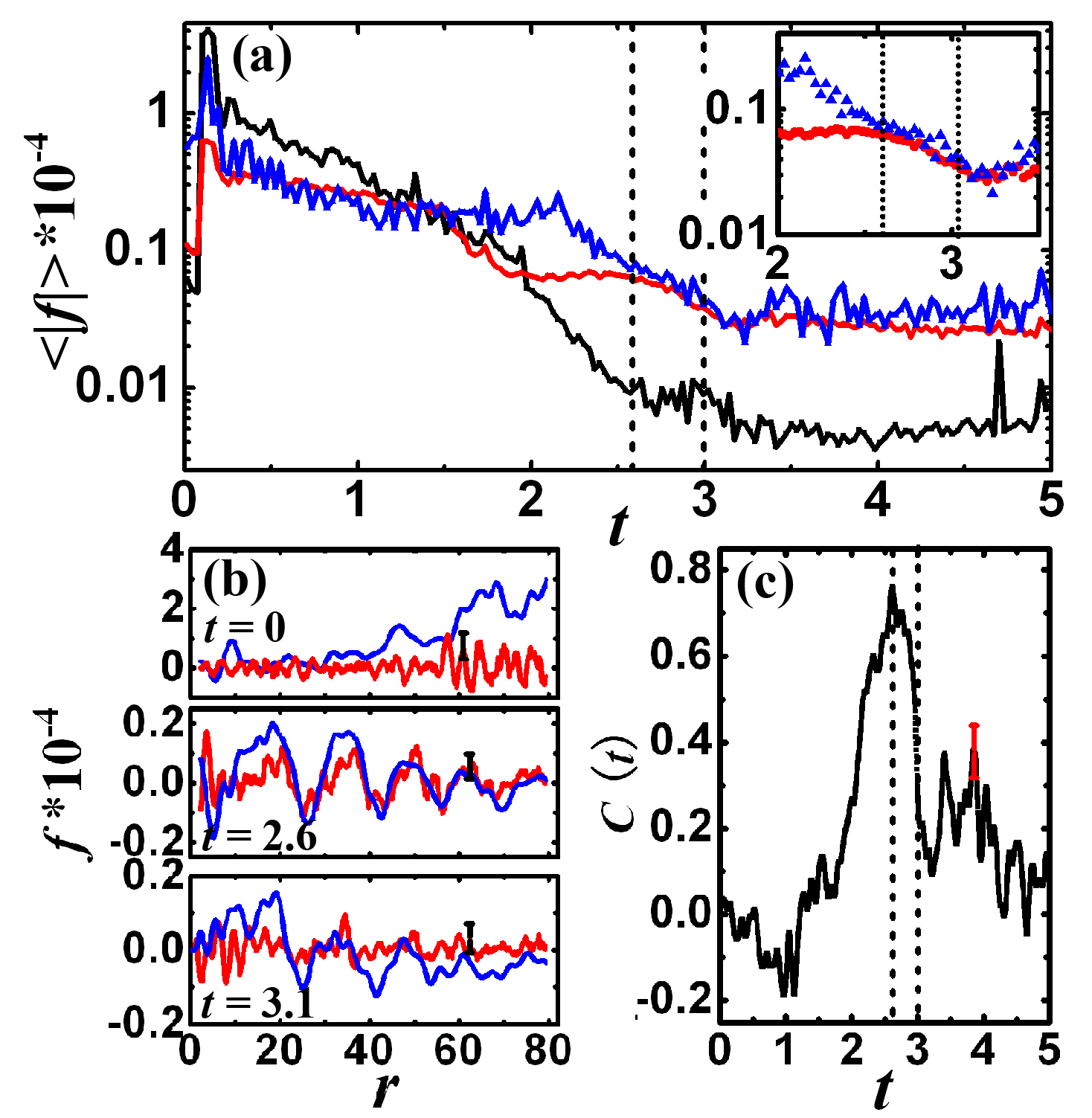}
\caption{Fluid accelerations measured in the experiment with $Ro=1.4{\times}10^{-3}$. (a) Averaged accelerations as function of time. Black curve: ${\langle}|f_i|{\rangle}$; blue curve: ${\langle}|f_c|{\rangle}$; and red curve: ${\langle}|f_{\nu} |{\rangle}$. The vertical dashed lines delineate when the ringed structure is maximal, $t=t_1=2.6$,  and when the rings started to break down $t=t_2=3.1$. The inset focuses on $t_1\le t \le t_2$.  (b) Fluid accelerations as functions of the radial distance at three stages. Blue curve: $f_c$; red curve: $f_{\nu}$. (c) Correlation function $C(t)$ between $f_c$ and $f_{\nu}$ as a function of time. Vertical bars denote fluctuation amplitudes.}
\label{Figure3}
\end{figure}

Because the flow is rotationally dominated we complete our analysis by considering the 
azimuthal component  of the momentum equation in a co-rotating reference frame in cylindrical-polar coordinates,  
\be
\frac{\partial{u_{\theta}}}{\partial{t}}+\left[(\vec{u} \cdot {\nabla}) \vec{u}\right]_{\theta}=\left[\nu{\nabla}^2\vec{u}\right]_{\theta}   -2 \left[\Omega{\times}\vec{u}\right]_{\theta},
\label{momentum}
\ee
where $\vec{u} = (u_r, u_{\theta})$ describes the radial and azimuthal velocity components and $\left[\vec{A}\right]_{\theta} $ denotes the azimuthal component of $\vec{A}$.  We nondimensionalize using the Ekman length $L_E=\sqrt{\nu/{\Omega}}$, and $\tau_E$ to extract  the Coriolis ($f_c$), viscous ($f_{\nu}$) and inertial ($f_i$) forces per unit mass from the PIV velocity fields by integration of each term of Eq. (\ref{momentum}) over $\theta = (0, 2\pi)$.  The areal averaged forces per unit mass for a component $f_j$ are defined as
\be
 {\langle}|f_j|{\rangle}=\frac{2}{r_{m}^2}{\int}^{r = r_{m}}_{r =0}|{f_j}|rdr, 
 \label{components}
 \ee
from the ring center ($r =0$)  to the maximum radius ($r = r_{m}$).   

The time evolution of ${\langle}|f_c|{\rangle}$, ${\langle}|f_{\nu} |{\rangle}$ and ${\langle}|f_i|{\rangle}$ (Fig.(\ref{Figure3})a) shows that upon spin-up the fluid inertia ${\langle}|f_i|{\rangle}$ decreases most rapidly and ${\langle}|f_c|{\rangle}$ converges with ${\langle}|f_{\nu} |{\rangle}$  as the ring state begins to appear at the center. 
Most striking is the abrupt appearance of the ring pattern which fills the cell surface at $t=t_1$  and then breaks down at $t=t_2$ (vertical dashed lines). Hence, the dissection of the velocity field shows that the origin of the transient ring state is that during $t_1\le t \le t_2$, ${\langle}|f_c|{\rangle}$ = ${\langle}|f_{\nu} |{\rangle}\sim 10 {\langle}|f_i|{\rangle}$; the Coriolis force balances the viscous force both of which dominate inertia.

Figs.(\ref{Figure3})b,c display the radial dependence of $f_c$ and $f_{\nu}$ at three times and their correlation function
\be
C(t)\equiv\frac{{\int}^{r = r_{m}}_{r =0}(f_c-{\langle}f_c{\rangle}){}(f_v-{\langle}f_v{\rangle})rdr}{\sqrt{\left[{\int}^{r = r_{m}}_{r =0}(f_c-{\langle}f_c{\rangle})^2rdr \right]{} \left[{\int}^{r = r_{m}}_{r = 0}(f_v-{\langle}f_v{\rangle})^2rdr\right]}} ,
\label{C(t)}
\ee
as a function of time, where the ${\langle}f_j{\rangle}$ are defined by Eq. (\ref{components}) but without taking the absolute value.  Thus, the maximal ring pattern of Fig.(\ref{Figure1})b appears as a maximum in $C(t)$ at $t=t_1$ when $f_c$ = $f_{\nu}$ {\em and} they are in phase as seen in the middle panel of Fig.(\ref{Figure3})b.  That the ring pattern breaks down at $t=t_2$ through a Kelvin-Helmholtz instability is demonstrated by the abrupt drop in $C(t)$ signaling the loss of phase between $f_c$ and $f_{\nu}$ at $t$ = 3.1 and the beginning of rotational dominance in the horizontal force balance driving shear and vortex formation.

\begin{figure}
\includegraphics[width=3.15in]{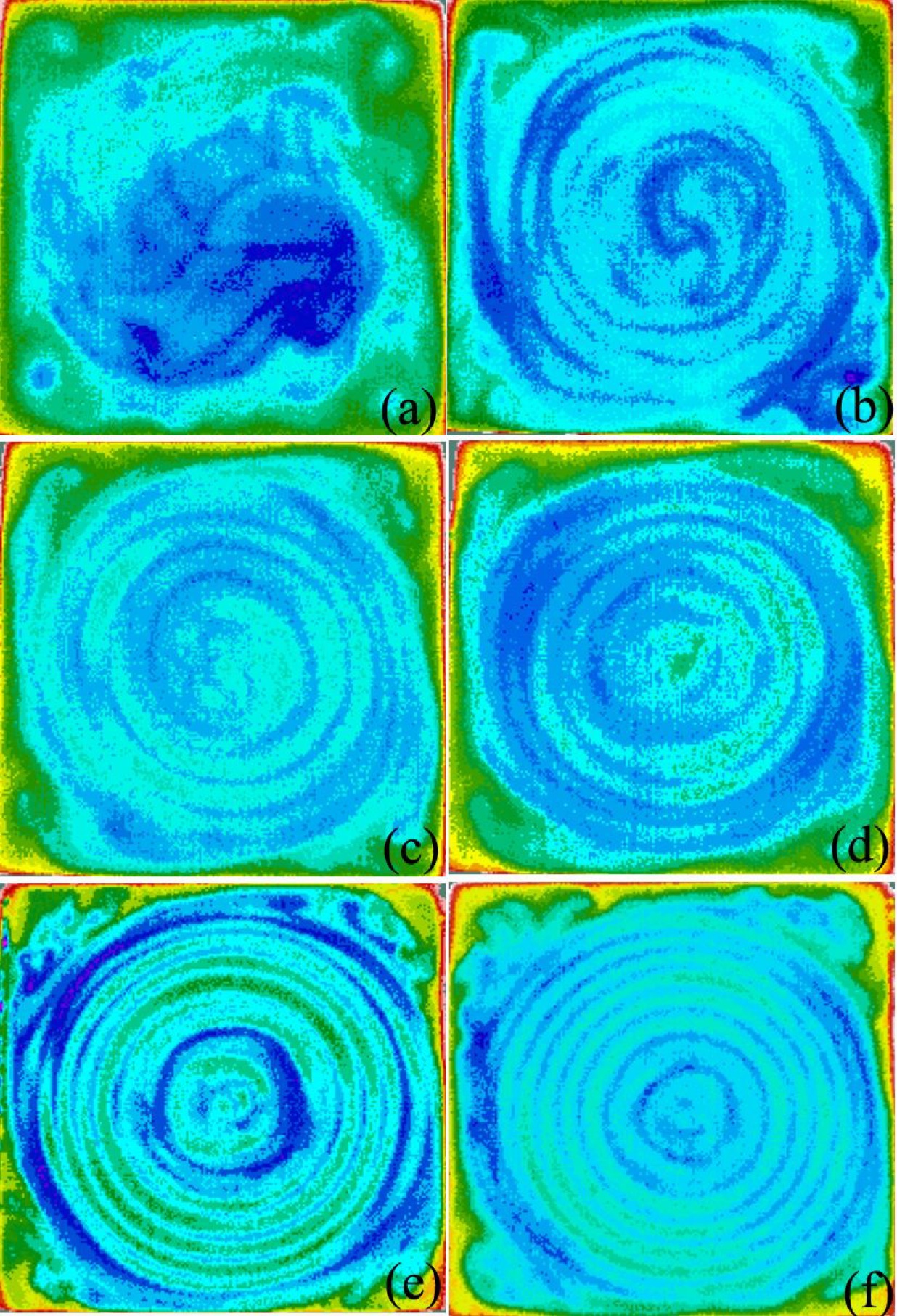}
\caption{Flow structures revealed by the thermal images for different $Ro$. From (a) to (f) $Ro$=$1.2{\times}10^{-2}$, $4.0{\times}10^{-3}$, $2.2{\times}10^{-3}$, $1.4{\times}10^{-3}$, $7.6{\times}10^{-4}$ and $5.0{\times}10^{-4}$ respectively. With the exception of (a) the image was taken when the ringed structure was maximal. Image (b) is taken at the threshold value $Ro_c$ for ring formation. }
\label{Figure4}
\end{figure}

The temperature maps in Fig. (\ref{Figure4}) show the flow pattern at different Rossby numbers when the ringed structure is maximal.  The ring wavelength ${\lambda}$ decreases with increasing ${\Omega}$  and there is a clear threshold Rossby number, here $Ro_c = 4.0{\times}10^{-3}$, above which the ring patterns are absent (Fig. \ref{Figure4}a). Indeed, for small ${\Omega}$ horizontal motion is too weak to suppress  vertical thermal convection; the flow resembles the ${\Omega}$ = 0 case of Fig. (\ref{Figure2})a.  When $t{\gg}t_2$ the ringed pattern evolves into a regular vortex grid.  After spin up we systematically varied $\Omega$ and measured the number of vortices per unit area $N$, and the mean $\lambda$ at $t=t_1$.  The average spacing between the rings (Fig.\ref{Figure1}b) is related to the mean distance between adjacent vortex cores (Fig.\ref{Figure1}d) according to ${\lambda} = C/\sqrt{N}$, with a best fit of $C = 0.605\pm0.015$.

\begin{figure}
\includegraphics[width=3.15in]{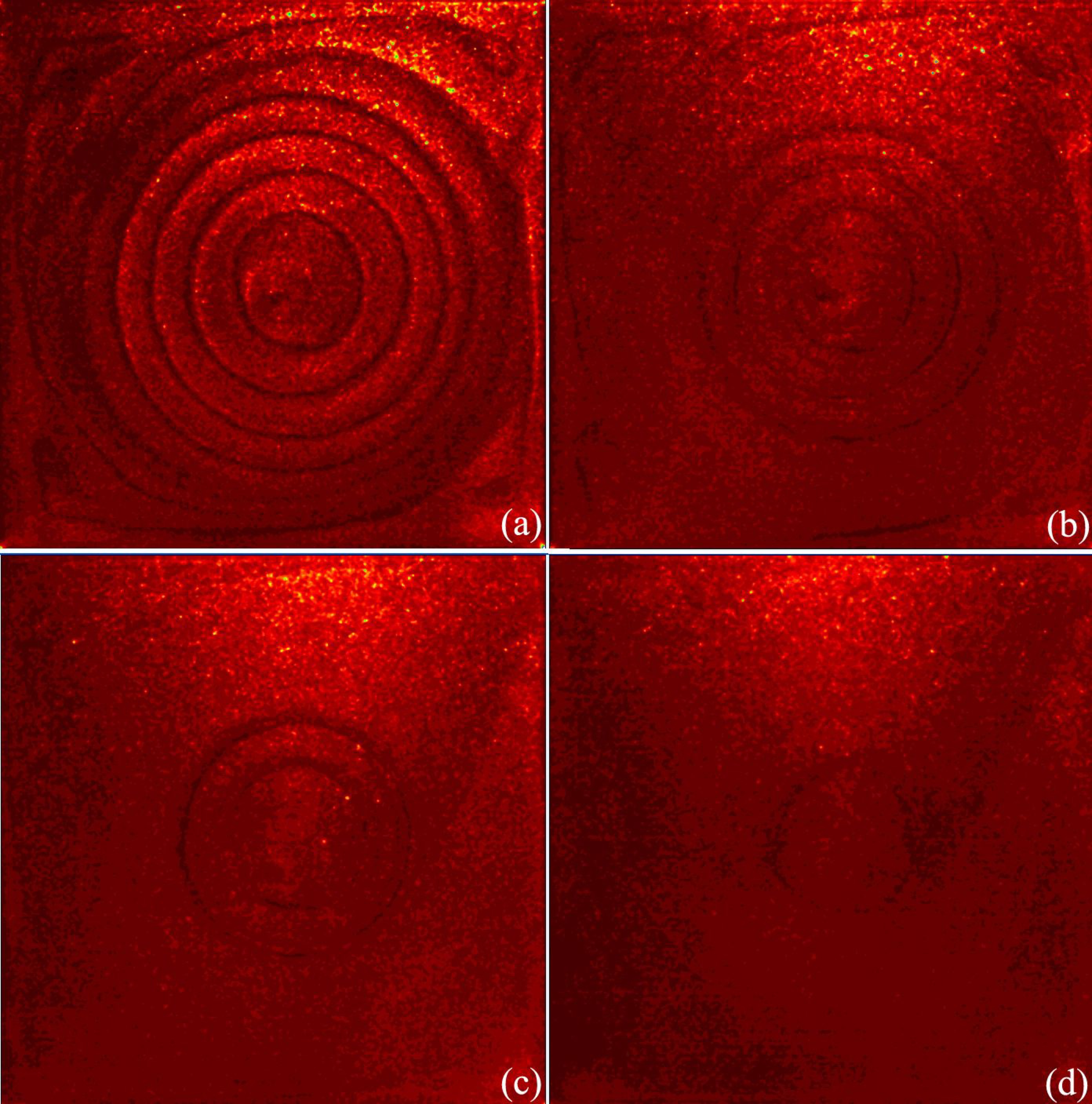}
\caption{Particle-number-density images versus fluid depth $z$ in cm. (a) $z=0.6$, (b) $z=3.1$, (c) $z=5.7$ and (d) $z=8.3$. Here $Ro=1.4{\times}10^{-3}$ the images are taken at $t=t_2$ when ring breakdown begins. }
\label{Figure5}
\end{figure}

Finally, to demonstrate the vertical extent of the ringed state we present the flow pattern simultaneously at different horizontal levels at the $t=t_2$ just when the surface pattern begins to break down (Fig. \ref{Figure5}).  Here we see the ``bowl'' structure of the ringed pattern with the central ring disappearing beyond a depth of $z=\frac{3}{4}H$ below the surface (Fig. \ref{Figure5}d).  Because of the finite lifetime of the rings, forming at the center and moving radially outward, the depth of fluid penetration necessarily decreases with radius.  

In summary we have used a combination of PIV and infrared thermometry to quantify the evolution of an evaporatively driven convecting spin-up system.  We find that, beginning at 2.6  $ \tau_E $, a balance of Coriolis and viscous forces dominating over inertia leads to a
transient concentric ringed convection pattern with radial sheets of downwelling and upwelling fluid.  When these dominant forces lose phase coherence, the sheets breakdown into the  vortex grid typical of quasi-geostrophic turbulence.  Moreover, the ring spacing and the vortex core spacing after ring breakdown obey a simple scaling relation.  Our methodology demonstrates the generality of the ringed pattern state.  As a practical matter, the great sensitivity found here of convection driven by evaporation under room temperature conditions suggests that many experiments with an exposed upper surface will display unintended dynamical effects.  Finally, we determined the critical Rossby number for the existence of a ringed state and this suggests possible implications for both astrophysical and geophysical convection wherein rotating convection is ubiquitous.

{\bf Acknowledgements}:
The authors gratefully acknowledge discussions and advice from A.J. Wells 
and support from the US National Science Foundation Grant No. OPP0440841, Helmholtz Gemeinschaft Alliance ``Planetary Evolution and
Life,'' and Yale University.


\begin{thebibliography}{19}
\expandafter\ifx\csname natexlab\endcsname\relax\def\natexlab#1{#1}\fi
\expandafter\ifx\csname bibnamefont\endcsname\relax
  \def\bibnamefont#1{#1}\fi
\expandafter\ifx\csname bibfnamefont\endcsname\relax
  \def\bibfnamefont#1{#1}\fi
\expandafter\ifx\csname citenamefont\endcsname\relax
  \def\citenamefont#1{#1}\fi
\expandafter\ifx\csname url\endcsname\relax
  \def\url#1{\texttt{#1}}\fi
\expandafter\ifx\csname urlprefix\endcsname\relax\def\urlprefix{URL }\fi
\providecommand{\bibinfo}[2]{#2}
\providecommand{\eprint}[2][]{\url{#2}}

\bibitem[{\citenamefont{Bondi and Lyttleton}(1948)}]{Bondi:48}
\bibinfo{author}{\bibfnamefont{H.}~\bibnamefont{Bondi}} \bibnamefont{and}
  \bibinfo{author}{\bibfnamefont{R.~A.} \bibnamefont{Lyttleton}},
  \bibinfo{journal}{Proc. Camb. Phil. Soc.} \textbf{\bibinfo{volume}{44}},
  \bibinfo{pages}{345} (\bibinfo{year}{1948}).

\bibitem[{\citenamefont{Greenspan and Howard}(1963)}]{Harvey:63}
\bibinfo{author}{\bibfnamefont{H.~P.} \bibnamefont{Greenspan}}
  \bibnamefont{and} \bibinfo{author}{\bibfnamefont{L.~N.}
  \bibnamefont{Howard}}, \bibinfo{journal}{J. Fluid Mech.}
  \textbf{\bibinfo{volume}{17}}, \bibinfo{pages}{385} (\bibinfo{year}{1963}).

\bibitem[{\citenamefont{Bretherton and Spiegel}(1968)}]{EAS:1968}
\bibinfo{author}{\bibfnamefont{F.~P.} \bibnamefont{Bretherton}}
  \bibnamefont{and} \bibinfo{author}{\bibfnamefont{E.~A.}
  \bibnamefont{Spiegel}}, \bibinfo{journal}{Astrophys. J.}
  \textbf{\bibinfo{volume}{153}}, \bibinfo{pages}{L77} (\bibinfo{year}{1968}).

\bibitem[{\citenamefont{Veronis}(1970)}]{GV:70}
\bibinfo{author}{\bibfnamefont{G.}~\bibnamefont{Veronis}},
  \bibinfo{journal}{Ann. Rev. Fluid Mech.} \textbf{\bibinfo{volume}{2}},
  \bibinfo{pages}{37} (\bibinfo{year}{1970}).

\bibitem[{\citenamefont{Duck and Foster}(2001)}]{Duck:2001}
\bibinfo{author}{\bibfnamefont{P.~W.} \bibnamefont{Duck}} \bibnamefont{and}
  \bibinfo{author}{\bibfnamefont{M.~R.} \bibnamefont{Foster}},
  \bibinfo{journal}{Ann. Rev. Fluid Mech.} \textbf{\bibinfo{volume}{33}},
  \bibinfo{pages}{231} (\bibinfo{year}{2001}).

\bibitem[{\citenamefont{Wells et~al.}(2007)\citenamefont{Wells, Clercx, and
  Van~Heijst}}]{WCV07}
\bibinfo{author}{\bibfnamefont{M.~G.} \bibnamefont{Wells}},
  \bibinfo{author}{\bibfnamefont{H.~J.~H.} \bibnamefont{Clercx}},
  \bibnamefont{and} \bibinfo{author}{\bibfnamefont{G.~J.~F.}
  \bibnamefont{Van~Heijst}}, \bibinfo{journal}{J. Fluid Mech.}
  \textbf{\bibinfo{volume}{573}}, \bibinfo{pages}{339} (\bibinfo{year}{2007}).

\bibitem[{\citenamefont{Flor et~al.}(2004)\citenamefont{Flor, Bush, and
  Ungarish}}]{Flor:2004}
\bibinfo{author}{\bibfnamefont{J.~B.} \bibnamefont{Flor}},
  \bibinfo{author}{\bibfnamefont{J.~W.~M.} \bibnamefont{Bush}},
  \bibnamefont{and} \bibinfo{author}{\bibfnamefont{M.}~\bibnamefont{Ungarish}},
  \bibinfo{journal}{Geophys. Astrophys. Fluid Dyn.}
  \textbf{\bibinfo{volume}{98}}, \bibinfo{pages}{277} (\bibinfo{year}{2004}).

\bibitem[{\citenamefont{Chandrasekhar}(1961)}]{Chandrabook}
\bibinfo{author}{\bibfnamefont{S.}~\bibnamefont{Chandrasekhar}},
  \emph{\bibinfo{title}{Hydrodynamic and Hydromagnetic Stability}}
  (\bibinfo{publisher}{Clarendon}, \bibinfo{year}{1961}).

\bibitem[{\citenamefont{Boubnov and Golitsyn}(1986)}]{BG:1986}
\bibinfo{author}{\bibfnamefont{B.~M.} \bibnamefont{Boubnov}} \bibnamefont{and}
  \bibinfo{author}{\bibfnamefont{G.~S.} \bibnamefont{Golitsyn}},
  \bibinfo{journal}{J. Fluid Mech.} \textbf{\bibinfo{volume}{167}},
  \bibinfo{pages}{503} (\bibinfo{year}{1986}).

\bibitem[{\citenamefont{Vorobieff and Ecke}(1998)}]{VE:1998}
\bibinfo{author}{\bibfnamefont{P.}~\bibnamefont{Vorobieff}} \bibnamefont{and}
  \bibinfo{author}{\bibfnamefont{R.~E.} \bibnamefont{Ecke}},
  \bibinfo{journal}{Phys. Fluids} \textbf{\bibinfo{volume}{10}},
  \bibinfo{pages}{2525} (\bibinfo{year}{1998}).

\bibitem[{\citenamefont{Patterson and Wettlaufer}(2010)}]{PIV}
\bibinfo{author}{\bibfnamefont{M.~D.} \bibnamefont{Patterson}}
  \bibnamefont{and} \bibinfo{author}{\bibfnamefont{J.~S.}
  \bibnamefont{Wettlaufer}}, \bibinfo{journal}{Rev. Sci. Instrum. (subjudice)}
  (\bibinfo{year}{2010}).

\bibitem[{\citenamefont{Berg et~al.}(1966)\citenamefont{Berg, Boudart, and
  Acrivos}}]{BERG:1966}
\bibinfo{author}{\bibfnamefont{J.~C.} \bibnamefont{Berg}},
  \bibinfo{author}{\bibfnamefont{M.}~\bibnamefont{Boudart}}, \bibnamefont{and}
  \bibinfo{author}{\bibfnamefont{A.}~\bibnamefont{Acrivos}},
  \bibinfo{journal}{J. Fluid Mech.} \textbf{\bibinfo{volume}{24}},
  \bibinfo{pages}{721} (\bibinfo{year}{1966}).

\bibitem[{\citenamefont{Boubnov and Golitsyn}(1990)}]{BG:1990}
\bibinfo{author}{\bibfnamefont{B.~M.} \bibnamefont{Boubnov}} \bibnamefont{and}
  \bibinfo{author}{\bibfnamefont{G.~S.} \bibnamefont{Golitsyn}},
  \bibinfo{journal}{J. Fluid Mech.} \textbf{\bibinfo{volume}{219}},
  \bibinfo{pages}{215} (\bibinfo{year}{1990}).

\bibitem[{\citenamefont{Marshall and Schott}(1999)}]{Marshall:1999}
\bibinfo{author}{\bibfnamefont{J.}~\bibnamefont{Marshall}} \bibnamefont{and}
  \bibinfo{author}{\bibfnamefont{F.}~\bibnamefont{Schott}},
  \bibinfo{journal}{Rev. Geophys.} \textbf{\bibinfo{volume}{37}},
  \bibinfo{pages}{1} (\bibinfo{year}{1999}).

\bibitem[{\citenamefont{Vorobieff and Ecke}(2002)}]{VE:2002}
\bibinfo{author}{\bibfnamefont{P.}~\bibnamefont{Vorobieff}} \bibnamefont{and}
  \bibinfo{author}{\bibfnamefont{R.~E.} \bibnamefont{Ecke}},
  \bibinfo{journal}{J. Fluid Mech.} \textbf{\bibinfo{volume}{458}},
  \bibinfo{pages}{191} (\bibinfo{year}{2002}).

\bibitem[{\citenamefont{Sprague et~al.}(2006)\citenamefont{Sprague, Julien,
  Knobloch, and Werne}}]{Sprague:2006}
\bibinfo{author}{\bibfnamefont{M.}~\bibnamefont{Sprague}},
  \bibinfo{author}{\bibfnamefont{K.}~\bibnamefont{Julien}},
  \bibinfo{author}{\bibfnamefont{E.}~\bibnamefont{Knobloch}}, \bibnamefont{and}
  \bibinfo{author}{\bibfnamefont{J.}~\bibnamefont{Werne}}, \bibinfo{journal}{J.
  Fluid Mech.} \textbf{\bibinfo{volume}{551}}, \bibinfo{pages}{141}
  (\bibinfo{year}{2006}).

\bibitem[{\citenamefont{Wells and Wettlaufer}(2008)}]{WW:2008}
\bibinfo{author}{\bibfnamefont{M.~G.} \bibnamefont{Wells}} \bibnamefont{and}
  \bibinfo{author}{\bibfnamefont{J.~S.} \bibnamefont{Wettlaufer}},
  \bibinfo{journal}{Geophys. Res. Lett.} \textbf{\bibinfo{volume}{35}},
  \bibinfo{pages}{L03501} (\bibinfo{year}{2008}).

\bibitem[{\citenamefont{Shang et~al.}(2003)\citenamefont{Shang, Qiu, Tong, and
  Xia}}]{Shang:2003}
\bibinfo{author}{\bibfnamefont{X.~D.} \bibnamefont{Shang}},
  \bibinfo{author}{\bibfnamefont{X.~L.} \bibnamefont{Qiu}},
  \bibinfo{author}{\bibfnamefont{P.}~\bibnamefont{Tong}}, \bibnamefont{and}
  \bibinfo{author}{\bibfnamefont{K.~Q.} \bibnamefont{Xia}},
  \bibinfo{journal}{Phys. Rev. Lett.} \textbf{\bibinfo{volume}{90}},
  \bibinfo{pages}{074501} (\bibinfo{year}{2003}).

\bibitem[{\citenamefont{Brown et~al.}(2005)\citenamefont{Brown, Nikolaenko, and
  Ahlers}}]{Brown:2005}
\bibinfo{author}{\bibfnamefont{E.}~\bibnamefont{Brown}},
  \bibinfo{author}{\bibfnamefont{A.}~\bibnamefont{Nikolaenko}},
  \bibnamefont{and} \bibinfo{author}{\bibfnamefont{G.}~\bibnamefont{Ahlers}},
  \bibinfo{journal}{Phys. Rev. Lett.} \textbf{\bibinfo{volume}{95}},
  \bibinfo{pages}{084503} (\bibinfo{year}{2005}).
  
  \bibitem[20]{Movie}
  \bibinfo{journal}{See EPAPS Document No. XX  for
the movie associated with the sequence in Fig. 1. For more information on
EPAPS, see http://www.aip.org/pubservs/epaps.html.}

\end{thebibliography}

\end{document}